# A review of Energy Efficient Routing Protocols in Underwater Internet of Things

Mehran Tarif
*Department of Computer Engineering,
Ardabil Branch, Islamic Azad
University,* Ardabil, Iran
0009-0003-0951-1800

Babak Nouri Moghadam*
*Department of Computer Engineering,
Ardabil Branch, Islamic Azad
University,* Ardabil, Iran
0000-0001-5363-9949

*Abstract*—Oceans, covering 70% of Earth's surface, are largely unexplored, with about 95% remaining a mystery. Underwater wireless communication is pivotal in various domains, such as real-time aquatic data collection, marine surveillance, disaster prevention, archaeological exploration, and environmental monitoring. The Internet of Things has opened new avenues in underwater exploration through the underwater Internet of Things concept. This innovative technology facilitates smart ocean research, from small case studies to large-scale operations. UIoT networks utilise underwater equipment and sensors to gather and transmit data in aquatic environments. However, the dynamic nature of these environments poses challenges to the network's structure and communication, necessitating efficient routing solutions. Quality-of-service-aware routing is vital as it minimises energy usage, extends battery life, and enhances network performance. This paper delves into the challenges and limitations of UIoT networks, highlighting recent routing methodologies. It also proposes a comparison framework for routing methods, focusing on the quality of service in underwater IoT networks, to foster more optimal route selection and better resource management.

I. INTRODUCTION

With the increasing growth of Internet of Things applications, researchers and related industries have noticed underwater sensor networks as an essential part of this technology in marine science. This field integrates telecommunication and computing platforms and is acknowledged as a segment of the broader UIoT issue. It encompasses both stationary and mobile sensors/actuators and underwater robots. [1]. Subsurface monitoring and monitoring in seas and oceans is critical due to their diverse environmental and industrial applications [2]. In the past, the focus was on the physical layer and signal processing, and less attention was paid to network discussion. However, due to extensive applications such as monitoring the underwater environment, it is necessary to expand the underwater network. Many sensors (such as sonar, optical, laser, magnetic, etc.) are placed underwater to enable monitoring [3]. Expanding this environment requires correct analysis of sensor output and their networking. Due to the high risk of the underwater ecosystem, the resource limitation in the underwater sensor network has more requirements to ensure its performance and stability [4].

IoT technology provides a comprehensive framework for underwater communications using various methods, including optical signals, magnetic induction, and sound. This technology can change industrial projects, scientific research, and business. One of the essential components to enable UIoT is an underwater wireless sensor network. However, currently, this technology faces challenges such as limited reliability, long propagation delay, high energy consumption, dynamic topology, and limited bandwidth. This study reviews the literature on inefficient energy routing in UIoT to pinpoint challenges and shortcomings and offers solutions to enhance service quality through improved energy efficiency [5]. The results of this study show that the essential elements available to meet future IoT challenges include underwater communications, energy storage, latency, mobility, lack of standardisation, transmission media, transmission range, and energy limitations.

The continuation of this article is divided into two parts. The architecture of the underwater Internet of Things network is discussed in the second part. The limitations and challenges of designing protocols in the underwater Internet of Things network structure with a look at their routing are given in the third section. The fourth section is devoted to routing in the underwater Internet of Things and their classification, and the fifth section discusses recent energy-efficient routing methods. Finally, the sixth section is devoted to the conclusion of the article.

II. UIoT ARCHITECTURE

The architecture and structure of the underwater Internet of Things network are categorised based on their entities and how they interact and cooperate. Some of the most critical categories of underwater Internet of Things networks include [6], [7]:

- 1D-UIoT network architecture: One-dimensional architecture includes sensors that receive, process, and send information to the base station. For example, floats are used to detect water features and can collect data for a certain period. This information is then transferred to the back of the float to send to the base station.
- 2D-UIoT network architecture: The 2D architecture consists of a set of sensor nodes (clusters) placed underwater. Each cluster contains a cluster head (an anchor node). Clusters are permanently located on the surface of the water. Each cluster member collects information and transmits it to the anchor node. The anchor node collects this information from all cluster member nodes and sends it to surface floating nodes. Communication in a 2D-UIOT network occurs in two dimensions: first, each cluster member interacts with its anchor node through horizontal communication, and second, the anchor node interacts with surface floating nodes through vertical communication links.
- 3D-UIoT network architecture: The 3D architecture consists of clusters at different ocean depths. This scenario is an extension of the 2D-UIoT architecture. Here, nodes within each cluster communicate with each other, cluster heads communicate with each other, and finally, cluster heads transmit information to the communication layer.
- 4D-UIoT network architecture: The 4D architecture includes underwater vehicles (ROVs) and clusters located at different ocean depths, like the 3D-UIoT architecture. In this mode, the ROV collects information from the cluster heads and then relays it to the floating object or directly to them, depending on the situation.

### III. CHALLENGES OF UIoT PROTOCOL DESIGN

Routing protocols in computer and telecommunication networks are a vital and influential part of network performance. In underwater sensor networks, location-based and location-independent routing protocols are generally divided. However, it should be noted that the water flow and the movement of marine creatures happen randomly. For this reason, location-based routing protocols are unsuitable for the underwater environment. In addition, using the Global Positioning System (GPS) is inefficient in the underwater environment [8]. In underwater wireless sensor networks, batteries usually power the nodes, and it is not possible to recharge the batteries. Therefore, routing protocols should be optimal in terms of energy consumption so that they can communicate and transmit data without errors and manage energy consumption properly.

Also, when the sensors collect the required information, this information must be sent to the base station on the water's surface. The transmission of information from sensor nodes to the base station is expensive. For this reason, energy consumption is a critical factor in designing routing protocols for underwater wireless sensor networks [9]. In designing these protocols, attention should be paid to the many limitations in the underwater environment. Among the critical issues needed in the design of routing protocols for underwater sensor networks and UIoT platforms, the following can be mentioned:

- Energy limitation: In underwater sensor networks, energy is a significant limitation. Batteries are not solar-charged or easily replaceable, so routing protocols must consider energy consumption a critical challenge to optimise battery life and avoid project failure.
- Load balance: In underwater sensor networks, load balance is vital. Routing protocols should distribute network resources reasonably and rationally to prevent weaknesses and instability. In case of problems such as Bottlenecks or Hotspots, the routing protocol should quickly and optimally find the best solution.
- Underwater Location: In an underwater sensor network, there is no problem with the GPS location information of nodes and their neighbours. Spatial information has limited access in this environment, and using GPS in shallow waters is associated with significant errors. Routing protocols must optimise routing tables using information such as node depth and neighbour list and manage issues such as round tripping and packet circulation in the network.
- Mobility of nodes and instability of the fluid environment: In the underwater environment, the sensor nodes are unstable and affected by water currents, contact with underwater organisms, and fluid waves. This mobility of nodes makes routing protocols have to deal with fundamental changes in the neighbour list, routing, route discovery and repair.
- Lack of failure detection system: There are no failure detection or configuration problems before data recovery and aggregation in the underwater network. This issue can lead to the complete failure of the monitoring project. Routing protocols must be able to detect and correct problems quickly to avoid this issue.
- Lack of real-time monitoring: In the underwater sensor network, the recorded data are not available until they are collected and processed at the base station, and several hours are delayed. This causes essential monitoring, such as earthquakes and tsunamis, to be delayed.
- Impossibility of real-time configuration of the system: In the underwater sensor network, the interaction between coastal control systems and real-time monitoring commands is impossible. This means system settings and configuration are not done in real-time, and responding quickly to certain events, such as natural disasters, is impossible.

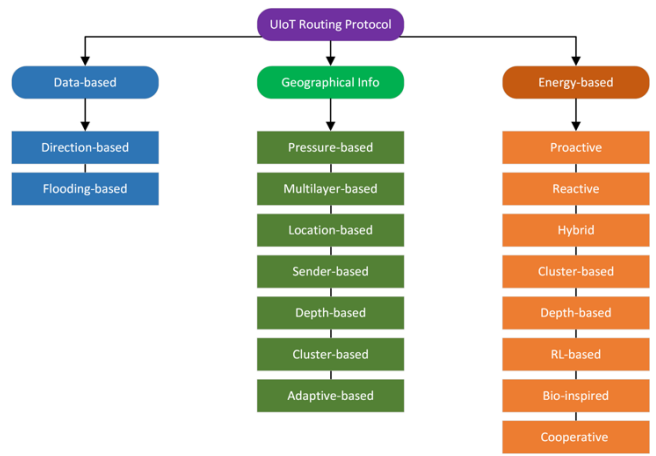

Fig. 1. Classification of routing protocols in underwater Internet of Things

### IV. ROUTING IN UNDERWATER INTERNET OF THINGS

Routing in the underwater Internet of Things network is designed based on various metrics [9]. This category is shown in Figure 1, and we have explained each category below.

#### A. Energy-based routing

The energy consumption of each node depends on the way of communication and signal processing load. The energy consumption of a node depends on three factors: distance, environmental factors, and battery capacity. In other words, the protocols that are in the energy-based category are implemented on one or more of the mentioned factors [10], [11].

- Reactive routing: When sending a packet, the reactive protocol initiates a path-finding approach to find the path to the destination. Reactive routing protocols are also known as on-demand routing protocols. They do not maintain a routing table. By doing so, they act as a bandwidth-efficient routing mechanism.
- Active routing: Each node in a network has one or more tables that describe the overall topology of the network. These tables are updated regularly to ensure each node has the most up-to-date routing information.

Topology information must be frequently sent between nodes to keep the routing information current, resulting in significant network overhead. On the other hand, on-demand routes will be available at any time.
- Hybrid routing is created from active and reactive routing and performs better than separately. Protocols divide the network into two regions and use separate mechanisms to communicate within the region and other regions. Somehow, depending on the needs of the network, they use one of the two methods interactively with other parts of the network.
- Cluster-based routing is one of the most popular routing techniques for wireless sensor networks. Nodes are grouped into a cluster, each with its cluster head. Each cluster head collects and transmits data from member nodes. The cluster head consumes more energy than the others when it broadcasts and collects information from the node. Therefore, the choice of the cluster head should be determined intelligently and according to the network conditions [12].
- Depth-based routing: In UIoT, the depth-based routing method is prevalent. The selection of the next node in depth-based routing depends on the depth of the node. The network topology follows the hierarchical method, and the forwarder selected by the origin is a node with a lower depth level than the forwarder. This indicates that the selected sender is closer to the sink node [13].
- RL-based routing: Human interaction with the world inspires reinforcement learning (RL). RL focuses on how intelligent agents behave in a complex environment. In RL, an agent achieves its goal by interacting and learning from its surroundings. RL learns about the environment, what to do, and how to condition the current behaviour to receive a numerical reward. The agent is rarely given instructions on which activities to perform and must rely on trial and error to learn which actions provide the best results. One of the most common RL approaches is Q learning. In learning Q, the agent makes a decision based on a specific Q value [14].
- Bio-inspired routing: In wireless networking, solving many heuristic problems is inspired by nature. These are natural solutions known as biologically inspired routing protocols. Some of the most well-known ones include ant colony, honey bee, frog jump, and so on [15].
- Reliability-based routing: One of the latest research topics in UIoT is cooperative-reliability routing, which considers the unstable underwater environment with the possibility of reliable data transmission from source to destination. In reliability-based routing, relay nodes forward packets from source to destination. Choosing the right relay nodes is the most critical part of these methods. The reliability-based routing approach helps to establish a reliable link between the source and the destination, thus increasing the throughput and packet delivery ratio. Destination nodes receive two or more copies of a packet, one from the source and one from the relay nodes. The sink node extracts the necessary information from several copies of the packet. However, the end-to-end delay increases and duplicate data transmission cannot be eliminated [16].

## B. Routing based on geographic information

Geographic routing, often known as location-based routing, is a simple and scalable solution [17], [18]. It does not require creating or maintaining entire routes to the destination. Additionally, no routing messages are required to update routing states. Instead, the path is chosen locally. At each hop, a local optimal next node, the nearest neighbour to the destination, is selected to continue sending the packet. This process is repeated until the package reaches its destination. Compared to packet retransmission, geographic routing can be used with opportunistic routing (OR) to increase data delivery and minimise energy consumption [8]. Each packet is forwarded to neighbours using an opportunistic routing paradigm. The priority of the nodes determines how the nodes are arranged.

Consequently, if a node's next step in the forwarding set successfully receives the packet, it forwards it only if its highest priority nodes do not. If a higher-priority node sends a packet, the next-priority node cancels the scheduled transmission of that packet. If all nodes in the set fail to receive a packet, opportunistic routing becomes the preferred mode for packet forwarding.

- Depth-based routing: In this method, only depth sensor data is needed to obtain the depth information of the node, and the complete information of the geographic location of the nodes is not needed. This method provides the right node and path selection based on node depth information in the next step. Also, network transmission delay and energy consumption are reduced in this method.
- Location-based routing: For this purpose, the sensor nodes must know the exact information of their geographic location. The best technique for route verification based on geolocation data is to create a route using node position data, such as angle and distance. Using geographic data, the source node selects the best neighbour node as the next step node. This method effectively avoids packet flooding and increases data transmission efficiency.
- Pressure-based routing: The levelling of sensors in the sea can be calculated using the barometer module because it is different at different water pressure levels. Greedy routing is one of the most popular and simple methods. One of the advantages of pressure-based sensors is that they only use position or depth information to transmit data and thus have no additional overhead.
- Sender-based routing: In sender-based routing protocols, the sending node selects its next step node alone. Depending on the type of application and the environment surrounding the network sensors, the criteria and metrics in choosing the next step are different depending on the point of view of the source node.
- Cluster-based routing: CBRP is a cluster-based routing protocol that works well in temporary applications and dynamic environments. This network divides the environment into different clusters, each with a cluster head. The head of the cluster is also responsible for tracking the membership in the cluster. Cluster membership information stored at each cluster head is used to discover inter-cluster paths dynamically. This technique effectively reduces traffic flooding during route discovery while simplifying the process by grouping nodes. In addition, it detects the presence of one-way links and uses them for intra-cluster and inter-cluster routing [12].

*C. Data-driven routing*

In this network, data is different depending on the type of application and often has different priorities. This sensor network has distinctions between information data in different areas. When an event occurs, the cluster of nodes in the area is notified and starts collecting data, which is subsequently sent to the sink for preparation. The data routing criterion in these networks is based on the priority and expiration time of packets [10].

- Direction-based addressing routing: In this method, the direction of the packet movement in the path is calculated and performed according to the default pattern. These direction-aware routing methods are directly related to data transmission efficiency.
- Flood routing: It is one of the oldest underwater routing methods. In this method, the node that receives the data information sends the packet to its neighbours until it reaches the destination node or exceeds a predetermined number of changes. Most of the routing protocols in this category are simple and efficient routing methods that do not require network topology or computational routing maintenance. Other advantages of these methods are simple structure and a high level of fault tolerance. However, there are issues such as internal message disclosure and resource wastage.

V. ENERGY-EFFICIENCY IN UIoT

Considering the trend of remote communication in the UIoT, multi-hop transmission is proposed as a high-priority method. However, the routing process to provide data communication faces several challenges, including delays, empty areas, and long propagation losses, which increase energy consumption [19].

Also, the traffic load and data interference affect the overall energy performance in the IoT network. To solve these problems, a clustering approach by grouping underwater nodes in clusters helps to reduce energy consumption. In this approach, each cluster has a cluster head for communication management [16], [20], [21].

Members of each cluster can send data to the cluster head to avoid interference and long transmission loss. Data aggregation falls under the responsibility of the cluster head to control data retransmission and save energy. Using the clustering approach of data traffic management enables the underwater Internet of Things network to remain active continuously and for a long time [22].

Both multi-hop routing protocols based on the clustering approach are the best choice for energy-efficient solutions for underwater IoT networks. However, the traditional clustering approach cannot provide an efficient and optimal solution due to various other issues, such as underwater fluctuations, high water pressure, propagation delay and low bandwidth [12]. Numerous studies have focused on enhancing optimal energy efficiency through smart and centralised clustering approaches to address these challenges. The primary emphasis of this research has been on the optimal selection of cluster heads and the decision-making process in routing.

Recently, much research has been proposed to design clustering protocols. However, most have neglected energy-aware trajectory planning and control in the UIoT network for the underwater operation of AUVs. The clustering approach should be improved by considering different aspects when choosing the number of clusters and cluster heads. These aspects include dynamic characteristics of underwater objects and other states related to nodes, such as position, energy, traffic load, distance from cluster head, and transmission strategy along inter-cluster or intra-cluster paths. These researchers have analysed various aspects of routing protocols based on developed approaches such as clustering, localisation topology, route planning and queue-based approaches [23]. The findings of these studies are concisely summarised in Table I.

In [24], Zhuo et al. introduced an underwater acoustic sensor network using AUV to solve the problems related to the energy limitations of underwater devices and the high demand for data collection. They have developed two efficient data collection algorithms using AUV. This study is presented to improve the performance of UWSN and balance energy consumption and network throughput. The maximum cluster problem is used to formulate the selection of cluster heads. They also present AUV path planning formulated as a travelling salesman problem to minimise the AUV travel time. However, this algorithm incurs delay due to additional computation because it uses a greedy method to find the near-optimal solution for cluster head selection.

Fang et al. [25] proposed a dataset using multiple AUVs in heterogeneous UIoT networks with optimisation (AoI). The proposed scheme uses a queuing model to enable information exchange between several AUVs. It also uses a low-complexity adaptive algorithm to control queue length and AUV energy constraints.

[26] have proposed an energy-efficient routing protocol for IoT networks. This protocol is adjusted based on the calculation of the residual energy of the nodes, the number of hops, the size of the data and the number of heterogeneous paths of the nodes. This scheme ensures that nodes with high residual energy can transmit packets and protect nodes with low energy from failure due to high traffic load. The evaluation of the protocol shows that it can significantly improve the energy efficiency of the IoT network and reduce the transmission delay.

In [27] the authors have presented a study that presents an algorithm to save energy in IoT networks by solving the problem of holes due to long routing distances between nodes. This algorithm uses a guard-based flooding scheme to handle the empty holes in the network. The evaluation results show that the proposed algorithm performs better than other well-known algorithms.

In [28] Bhattacharjya et al. present an efficient routing protocol for clustering-based underwater wireless sensor networks. This proposed protocol supports the monitoring and controlling of underwater scenarios in the field of IoT energy efficiency. It uses deploying network nodes with multi-hop communication, minimising energy consumption. The protocol calculation is based on the hop-to-hop communication technique, and the number of hops depends on the underwater relay nodes as independent and selected units. Information is transmitted to cluster nodes through underwater relay nodes. The designed protocol has been evaluated and shown to provide the lowest total energy consumption compared to AODV and other protocols.

TABLE I. RECENT ENERGY EFFICIENT ROUTING PROTOCOLS AND QUALITY OF SERVICE

| Protocol | Algorithm | Method | Performance indicators | | | |
|---|---|---|---|---|---|---|
| | | | Energy Efficiency | Lifetime | QoS | Efficiency |
| [24] 2020 | AEEDCO and AEEDCO-A are presented for power consumption in AUV | Optimising the performance of AUVs to maximise energy efficiency | ✓ | High | Low | Fair |
| [25] 2021 | AoI Optimisation in Heterogeneous IoT Network | Improved queuing model to enable information exchange between multiple AUVs | ✓ | Low | Low | Fair |
| [26] 2021 | Improved Energy Efficiency Routing Protocol | Balancing traffic load between nodes with higher residual energy | ✓ | Low | Low | Low |
| [27] 2021 | Guard-based flood pattern | It reduces energy consumption by solving the problem of holes due to the long distance of the route. | ✗ | High | Low | Low |
| [28] 2022 | Cluster approach based on energy-efficient routing protocol | Reducing energy consumption and the possibility of network development with multi-step communication. | ✓ | Low | Low | High |
| [29] 2022 | Opportunistic location-based routing protocols | Increasing energy efficiency and reducing latency. | ✓ | Low | High | Low |
| [13] 2022 | Step Selective Power Routing Protocol | Increase selection accuracy to reduce energy consumption and improve the delivery ratio. | ✓ | Low | Low | Justly |
| [30] 2022 | Combining Bayesian multidimensional scaling localisation | Increasing the accuracy of hybrid communication in underwater Internet of Things. | ✗ | High | Low | Justly |
| [31] 2022 | k-means algorithm | Optimal cluster head selection based on the distance to the base station | ✓ | High | Low | Justly |

Menon et al. have presented a study showing that location-based routing protocols use new mechanisms for underwater IoT devices with a clustering approach to improve energy efficiency. The authors have explained that most protocols improve quality of service or energy efficiency, but few benefit from both [29]. The authors also present a study to discuss the opportunistic routing protocol to increase energy efficiency in underwater IoT platforms and improve delivery time, as the analysis shows that this protocol provides better results than current protocols in terms of energy efficiency and quality of service.

Manal et al. reviewed Issues that can affect underwater communications performance. In addition, the authors propose a Directional Selective Power Routing (DSPR) protocol to reduce energy consumption and improve the delivery ratio. This proposed protocol selects the best path to the surface well based on the depth and angle of underwater entry [10], [13]. Also, the protocol can provide better connectivity by using a selective power control approach. The authors have stated that the proposed protocol outperforms fixed direction routing (DR) and variable depth-based routing (VDBR) protocols and reduces energy consumption by 30%.

Khalil et al. have proposed a hybrid Bayesian multidimensional scaling localisation technique that enables hybrid magnetic, optical, and audio communications for underwater IoT devices [30]. This technique uses a nearest-neighbour distance-based localisation approach for better underwater communication accuracy. The proposed scheme's evaluation shows that it performs better at the sub-meter accuracy level for underwater communication.

In [31], the protocol proposed by Li et al. is an energy-aware clustering protocol based on the k-means algorithm. This protocol influences the selection of cluster head candidates based on the location of the nodes and the amount of residual energy. The k-means algorithm helps select cluster heads according to their distance from the base station. In addition, this protocol uses a binary routing technique to reduce data transmission in clusters. The evaluation of this protocol shows that it effectively reduces the energy consumption and the number of dead nodes and provides better performance than the LEACH protocol.

VI. CONCLUSION

Underwater Internet of Things technology is one of the famous research topics today, which could change industrial projects and scientific and commercial research. An essential component to enable IoT is the underwater wireless sensor network. However, currently, this technology faces challenges such as limited reliability, long propagation delay, high energy consumption, dynamic topology, and limited bandwidth. This study conducted a literature review on the UIoT to pinpoint potential challenges and risks and to propose solutions to mitigate them. The results of this study show that the essential elements available to meet future IoT challenges include underwater communications, energy storage, latency, mobility, lack of standardisation, transmission media, transmission range, and energy limitations. In addition, in this study, an overview of recent methods and their comparison based on service quality criteria has been presented in the form of a table, which can pave the way for other researchers in this field.